\documentclass{article}

\usepackage{graphicx}
\usepackage{psfig}
\usepackage{epsfig}
\usepackage[round]{natbib}

\title{
Statistical modelling of tropical cyclone tracks: a comparison of models for the variance of trajectories
}

\author{Stephen Jewson}
\begin{document}

\author{Tim Hall, GISS\footnote{\emph{Correspondence address}: Email: \texttt{tmh1@columbia.edu}}\\and\\
Stephen Jewson\\}

\maketitle

\begin{abstract}
We describe results from the second stage of a project to build a statistical model for hurricane tracks.
In the first stage we modelled the unconditional mean track. We now attempt to model
the unconditional variance of fluctuations around the mean.
The variance models we describe use a semi-parametric nearest neighbours approach in which the
optimal averaging length-scale is estimated using a jack-knife out-of-sample fitting procedure.
We test three different models. These models
consider the variance structure of the deviations from the unconditional mean track to be isotropic,
anisotropic but uncorrelated, and anisotropic
and correlated, respectively. The results show that, of these models, the anisotropic correlated model
gives the best predictions of the distribution of future positions of hurricanes.
\end{abstract}

\section{Introduction}

We are interested in developing accurate methods to estimate the probability of extreme hurricanes
making landfall at different locations.
Estimating the probability of extreme events is difficult because,
by the definition of what makes an extreme, the data is sparse.
If we try and estimate hurricane landfall probabilities
using local data then this problem is acute: severe hurricanes only strike individual
geographical locations very rarely.
For some locations there are no strikes
at all in the historical record, even though such strikes may be possible.
However, the problem of lack of data can be reduced, and the risk estimated more accurately, by
appropriate use of data from surrounding regions. Various methods have been described that use this
principle to estimate hurricane risks, such as those of~\citet{clark86}, \citet{darling91} and~\citet{chu98}.
One of the most interesting approaches
is to build a statistical model for hurricane tracks and intensities
across the whole of the Atlantic basin.
This is the approach taken in~\citet{drayton00}, \citet{vickery00} and~\citet{emanuel05},
and is the approach we are taking ourselves.

In general, when using surrounding data to estimate local parameters,
one must decide on the size and shape of the region from which
data is to be taken and how the data within that region is to be weighted.
There is a balance between two effects: using more data gives more precise estimates,
but using less relevant data can introduce biases.
Finding where the optimal balance lies
is crucial to any attempt to build an accurate model.

In~\citet{hallj05a} we have started building a new model for hurricane tracks in the Atlantic basin,
and we address the question of
how best to use surrounding data by calculating the optimum size of the data region using
cross-validation.
The track model that we describe is based on a two dimensional linear stochastic process,
with the two dimensions representing
the longitude and latitude of the location of a hurricane.
In~\citet{hallj05a}, and also in this article, we assume that the innovations driving this stochastic process are Gaussian.
Because of this simplifying assumption the model can be completely specified by
the appropriate means, variances and covariances.
As a start, in~\citet{hallj05a} we describe a model for the unconditional mean.
That is, a model for the expected motion of a hurricane, given no information at all except the current
location of the hurricane.
The model uses a nearest neighbours approach with an isotropic Gaussian weighting function incorporating
a free parameter length-scale.
The cross-validation fitting procedure gives an optimal length-scale of 300km.

In this article we describe what we consider to be the most appropriate next stage in
building our track model, which is the modelling of the unconditional
variance of fluctuations around the unconditional mean tracks.
What this adds to the model for the unconditional mean tracks can be described as follows.
The model for the unconditional mean can be used to make one step (6 hour) predictions for the future
location of a hurricane.
A model for the unconditional mean \emph{and} variance goes a step further and
gives one step predictions for the whole \emph{distribution}
of future locations of the hurricane.
Again, the meaning of `unconditional' is that this prediction assumes no information at all except for the current
location. We are
ignoring autocorrelations in time as well as other influences on the track
such as the effects of intensity, time of year and ENSO state. All of these will be included
in the model in due course.

We estimate the unconditional variance using a nearest neighbours
approach, very similar to the method used to estimate the unconditional mean.
Broadly speaking, this approach works
as follows. Using the model for the unconditional mean we
separate the observed hurricane tracks into an unconditional mean component
and a deviation from the unconditional mean.
We then model the variance of these deviations at each point in the basin
as being equal to an empirical variance estimated from the observed deviations
near to that point.
As before, a length-scale defines what we mean by `near'.
We calculate a new length-scale specifically for the variance to allow for the possibility
that the length-scale could be different from that derived for the mean.

Within this overall framework
we will compare three different models for the variance, ranging from simple to complex.
We take this systematic simple-to-complex approach to try to ensure that we avoid over-fitting
(i.e. we want to avoid using
a complex model which performs less well than a simpler model could).
The models we are test are
(a) one in which the variance structure of the deviations from the mean is isotropic,
(b) one in which the variance structure of the deviations is anisotropic but uncorrelated
relative to orthogonal axes defined along and across the local
unconditional mean trajectory, and
 (c) one in which the variance structure of the deviations is anisotropic and \emph{correlated} relative to these axes.

We now describe the data we will use for this study (section~\ref{data}),
the three models (section~\ref{models}),
and the results from a comparison between the results of the three models (section~\ref{results}).
We then summarise the results and describe our future plans (section~\ref{summary}).
Finally we include two appendices which contain some additional information and discussion.

\section{The data}
\label{data}

The basic data set for this study is a subset of the Hurdat data, as described in~\citet{hallj05a}.
In that study we used a nearest neighbours method to predict the unconditional mean
of our stochastic process.
We use that model to make one-step (6 hour) predictions of the observed hurricane
tracks. By comparing these predictions with the actual tracks we can generate forecast errors.
We take these forecast errors as the starting point for the current study.
We will refer to these forecast errors as the deviations from the unconditional mean track.

\section{The models}
\label{models}

\subsection{The isotropic model}

Our initial model for the deviations from the unconditional mean track
is isotropic in that we assume that the variance of the deviations
is the same in all directions.
In other words, the contours of constant probability density in the
predicted distribution of the deviations are circular.
In the continuous time limit we can write this model as:
\begin{eqnarray}\label{isotropic}
 dX&=&\mu_x(\theta,\phi) dt+\sigma(\theta,\phi)dW_x\\\nonumber
 dY&=&\mu_y(\theta,\phi) dt+\sigma(\theta,\phi)dW_y
\end{eqnarray}

where $X$ and $Y$ are the longitude and latitude of the hurricane, $\theta$ and $\phi$ are longitude
and latitude, $\mu_x$ and $\mu_y$ are the unconditional mean velocities
(determined by the model for the unconditional mean), $\sigma$ is the standard
deviation of the deviations from the unconditional mean track
and $dW_x$ and $dW_y$ are independent Brownian motions.
The standard deviation $\sigma(\theta,\phi)$ is determined by the lengthscale $\lambda$ and the historical
data, so we could write:
\begin{equation}
 \sigma(\theta,\phi)=\sigma(\mbox{historical data},\lambda)
\end{equation}

In order to fit $\sigma$ at the point $(\theta,\phi)$
we calculate the weighted variance of the observed deviations from
the unconditional mean track model, where
the deviations are weighted using a Gaussian weighting function so that
nearest errors are much more important than distant errors.
We vary the lengthscale in the weighting function to find which length-scale gives the optimal results.

How should we define `optimal' for a probabilistic prediction model of this type?
When fitting the model for the unconditional mean track, we used RMSE as the cost function.
However, RMSE cannot be used to fit the variance, since changing the predicted variance
doesn't affect it. The most obvious generalisation of RMSE
seems to be (minus one times) the log-likelihood of classical statistics,
and this is what we use as our cost function.
We calculate the log-likelihood out of sample: an in-sample maximisation of the log-likelihood would
lead to an optimal lengthscale of zero.
It would also not take into account parameter uncertainty when comparing the results from different models,
and would thus not penalise over-fitted models.

We now derive an expression for the log-likelihood for the isotropic model.

\subsubsection{Likelihood for the isotropic model}

The multivariate normal distribution with dimension $p$ has the density:
\begin{equation}
 f=\frac{1}{(2\pi)^{\frac{p}{2}} D^\frac{1}{2}} \mbox{exp}\left(-\frac{1}{2}(z-\mu)^T\Sigma^{-1}(z-\mu)\right)
\end{equation}
where
$\Sigma$ is the covariance matrix (size $p$ by $p$),
$D$ is the determinant of the covariance matrix (a single number),
$z$ is a vector length $p$ and
$\mu$ is a vector length $p$.

In our two dimensional ($p=2$) isotropic case we let
$z=(x,y)$ be the deviations from the unconditional mean track,
we define $\sigma^2$ to be the variance of both $x$ and $y$,
and we set $\mu=0$ because we are looking at deviations.
The assumption of isotropy means that $x$ and $y$ are independent,
and so the covariance matrix $\Sigma$ is just $\sigma^2$ times the unit matrix:

\begin{equation}
\Sigma=
    \left(
\begin{array}{cc}
  \sigma^2 & 0 \\
  0 & \sigma^2 \\
\end{array}
\right).
\end{equation}

It follows that the inverse of the covariance matrix is given by:

\begin{equation}
\Sigma^{-1}=
    \left(
\begin{array}{cc}
  \sigma^{-2} & 0 \\
  0 & \sigma^{-2} \\
\end{array}
\right)
\end{equation}

and
\begin{equation}
D=\sigma^4.
\end{equation}

This gives:
\begin{eqnarray}
(z-\mu)^T\Sigma^{-1}(z-\mu)
 &=& \frac{1}{\sigma^2}(x^2+y^2)
\end{eqnarray}


For $N$ data points the likelihood is thus:
\begin{equation}
 f=\prod_{i=1}^{N} \frac{1}{2\pi \sigma_i^2} \mbox{exp} \left( -\frac{x_i^2+y_i^2}{2 \sigma_i^2} \right)
\end{equation}

and the log of this is
\begin{eqnarray}
 \mbox{ln}f
         &=&\sum_{i=1}^{N} -\mbox{ln} (2\pi \sigma_i^2) +\sum -\frac{x_i^2+y_i^2}{2 \sigma_i^2}
\end{eqnarray}

\subsection{The anisotropic uncorrelated model}

Our second model for the deviations from the mean track is anisotropic in that we allow for the deviations to have
different variances in the different directions defined by the coordinate system.
Rather than use the $(\theta,\phi)$ coordinate system of lines of longitude and latitude,
it seems to make sense to use a coordinate system defined to suit the problem.
The obvious choice is to use an orthogonal curvilinear coordinate system based on the
local unconditional mean tracks, as defined in~\citet{hallj05a} (see figure 3 in that paper).
We write the deviations from the unconditional mean track within this coordinate system as $(u,v)$.
$u$ represents displacements along the mean track while $v$ represents displacements across the mean track.
Within this coordinate system we assume that the forecast errors are independent.
The contours of constant probability density in this model are thus ellipses with their principal axes lying
along and across the directions given by the unconditional mean tracks.
By comparing the fitted values of the variance in the $u$ and $v$ directions we will be able
to see which of the principle axes is the longer of the two.

We can write this model as:

\begin{eqnarray}
 dU&=&\mu_u(\theta,\phi) dt+\sigma_u(\theta,\phi)dW_u\\\nonumber
 dV&=&\mu_v(\theta,\phi) dt+\sigma_v(\theta,\phi)dW_v
\end{eqnarray}

where U and V are the projections of the hurricane motion in the directions parallel and perpendicular,
respectively, to the local mean track, and $dW_u$ and $dW_v$ are uncorrelated.

\subsubsection{Likelihood for the anisotropic uncorrelated model}

We define
$z=(u,v)$ (deviations from the mean track in the along-mean-track and across-mean-track directions),
and the variances of $u$ and $v$ to be $\sigma_u^2$ and $\sigma_v^2$.
$\mu=0$ because we are looking at deviations from the mean tracks,
and because we assume that $u$ and $v$ are independent
the covariance matrix $\Sigma$ is diagonal and given by:

\begin{equation}
\Sigma=
\left(
\begin{array}{cc}
  \sigma_u^2 & 0 \\
  0          & \sigma_v^2 \\
\end{array}%
\right).
\end{equation}

The inverse of the covariance matrix is:

\begin{equation}
\sigma^{-1}=
\frac{1}{\sigma_u^2 \sigma_v^2}
\left(
\begin{array}{cc}
  \sigma_v^2 & 0 \\
  0          & \sigma_u^2 \\
\end{array}%
\right)
\end{equation}

and
\begin{equation}
D=\sigma_u^2 \sigma_v^2.
\end{equation}

This gives:
\begin{eqnarray}
(z-\mu)^T\Sigma^{-1}(z-\mu)
 &=& \frac{1}{\sigma_u^2\sigma_v^2}(u^2 \sigma_v^2+v^2 \sigma_u^2)
\end{eqnarray}

and so the log-likelihood for $N$ data points is:

\begin{eqnarray}
 \mbox{ln}f
         &=&\sum_{i=1}^{N} -\mbox{ln} (2\pi \sigma_u \sigma_v) +\sum -\frac{1}{2\sigma_u^2\sigma_v^2}(u_i^2 \sigma_v^2+v_i^2 \sigma_u^2)
\end{eqnarray}

where we have suppressed the $i$ subscripts on $\sigma_u$ and $\sigma_v$ for clarity.

\subsection{The anisotropic correlated model}

Our third model for the deviations from the unconditional mean tracks is isotropic but correlated.
As with the previous model we use the $(u,v)$ coordinate system
based on the local unconditional mean tracks,
but now we also calculate the correlation between along-track and across-track
errors in this coordinate system.
The contours of constant probability in this model are now ellipses but with arbitrary alignment relative
to the mean tracks. This alignment is determined by the correlation between the deviations in the
$u$ and $v$ directions.

We can write this model as:

\begin{eqnarray}
 dU&=&\mu_u(\theta,\phi) dt+\sigma_u(\theta,\phi)dW_u\\\nonumber
 dV&=&\mu_v(\theta,\phi) dt+\sigma_v(\theta,\phi)dW_v
\end{eqnarray}

where $dW_u$ and $dW_v$ are correlated with linear correlation coefficient $r$.

\subsubsection{Likelihood for the anisotropic correlated model}

The covariance matrix $\Sigma$ is now given by

\begin{equation}
\Sigma
=
\left(
\begin{array}{cc}
  \sigma_u^2          & r \sigma_u \sigma_v \\
  r \sigma_u \sigma_v & \sigma_v^2          \\
\end{array}%
\right)
\end{equation}

The inverse of the covariance matrix is

\begin{equation}
\Sigma^{-1}
=
\frac{1}{(1-r^2) \sigma_u^2 \sigma_v^2}
\left(
\begin{array}{cc}
  \sigma_v^2            & -r \sigma_u \sigma_v \\
  -r \sigma_u \sigma_v  & \sigma_u^2 \\
\end{array}%
\right)
\end{equation}

and
\begin{equation}
D
=(1-r^2)\sigma_u^2 \sigma_v^2.
\end{equation}

This gives:
\begin{eqnarray}
(z-\mu)^T\Sigma^{-1}(z-\mu)
 &=& \frac{1}{(1-r^2)\sigma_u^2\sigma_v^2}(u^2 \sigma_v^2+v^2 \sigma_u^2-2ruv\sigma_u\sigma_v)
\end{eqnarray}


and so the log-likelihood for $N$ data points is:

\begin{eqnarray}
 \mbox{ln}f
         &=&\sum_{i=1}^{N} -\mbox{ln} [2\pi (1-r^2)^\frac{1}{2} \sigma_u \sigma_v]
           +\sum -\frac{1}{2(1-r^2)\sigma_u^2\sigma_v^2}(u_i^2 \sigma_v^2+v_i^2 \sigma_u^2-2ruv\sigma_u\sigma_v)
\end{eqnarray}

Again we've suppressed the $i$ subscripts on $\sigma_u$, $\sigma_v$, and $r$ for clarity.

\section{Results}
\label{results}

\subsection{Log-likelihood scoring results}

\begin{figure}[!htb]
  \begin{center}
    \scalebox{0.8}{\includegraphics{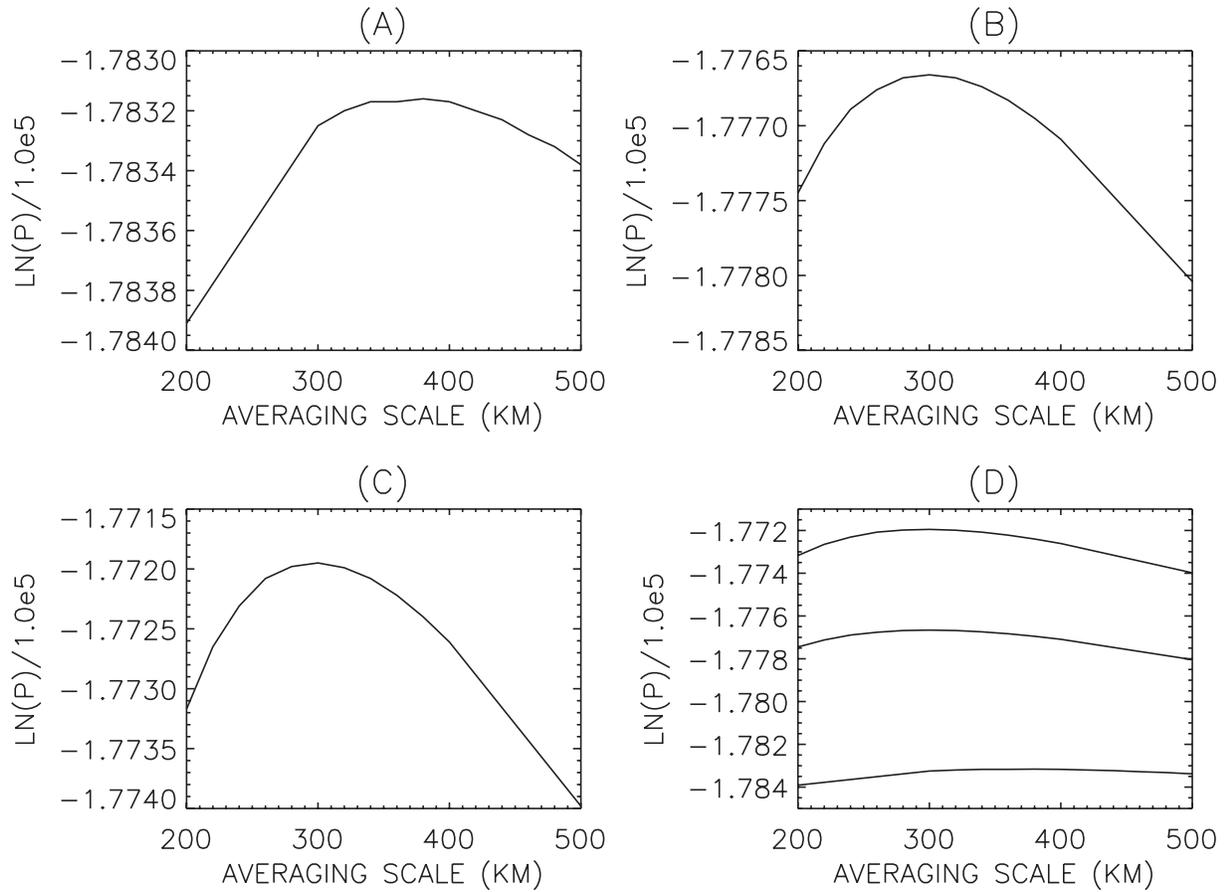}}
  \end{center}
  \caption{
Panels A, B and C show the log-likelihood score for the isotropic,
anisotropic and uncorrelated, and anisotropic and correlated models
discussed in the text, as a function of averaging length scale.
We see that the models have optimum averaging length-scales of
380km, 300km and 300km respectively.
Panels D shows the curves from panels A, B and C together.
The top curve (and hence the best model) is the anisotropic correlated
model. The middle curve is the anisotropic uncorrelated model and the lower
curve is the isotropic model.
     }
  \label{f01}
\end{figure}

Figure~\ref{f01} shows the log-likelihood score versus averaging length-scale for the three models.
For the isotropic model the optimal averaging length-scale is 380km (panel A).
For the anisotropic uncorrelated model the optimal averaging length-scale is 300km (panel B),
and for the anisotropic correlated model the optimal averaging length-scale is also 300km (panel C).
These optimal length-scales were calculated to within 20km.

\begin{figure}[!htb]
  \begin{center}
    \scalebox{0.8}{\includegraphics{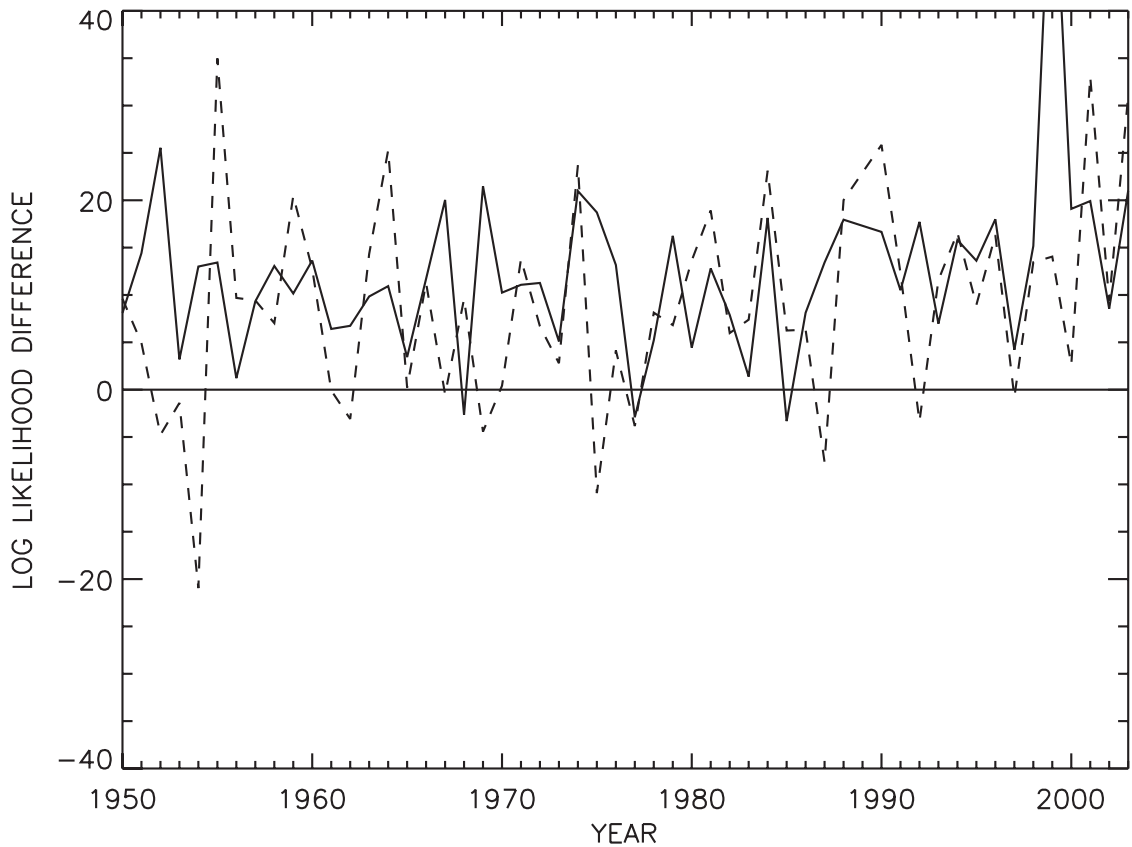}}
  \end{center}
  \caption{
Differences between the log-likelihood scores for the three variance
models on a year by year basis. The solid curve shows differences between
the two simplest models: the anisotropic uncorrelated model and the isotropic model. Since these
differences are almost all positive we conclude that the anisotropic uncorrelated
model beats the isotropic model in almost all years. The dashed line shows the
differences between the anisotropic correlated model and the anisotropic uncorrelated
model. Again the differences are mostly positive, and we can conclude that the
correlated model beats the uncorrelated model in most years.
          }
  \label{f02}
\end{figure}

In figure~\ref{f01}, panel D, we compare the log-likelihood scores
for the three models. We see that the
isotropic model gives the worst results, the anisotropic uncorrelated model
gives somewhat better results and the anisotropic correlated model gives the best of the three.
One could, on the basis of this comparison, conclude that the anisotropic correlated model
is the best model to use. However, it is possible that these differences are not really significant
i.e. that in fact the differences arise just because of the particular sample of data we are working
with, and that other samples would give a different ordering for the models. If this were the case then
it might be better to use the isotropic model on the basis that it is simpler to understand and
implement. It is therefore important to assess the significance of the differences in log-likelihood scores.
One way to assess whether the differences are significant
is to look at them for each individual year, and this is shown in figure~\ref{f02}.
The black curve shows the differences in the log-likelihood scores between the isotropic and
anisotropic uncorrelated models, and the dashed curve shows the differences between the anisotropic
uncorrelated model and the anisotropic correlated model. Comparing the isotropic and anisotropic
uncorrelated models we see that the anisotropic uncorrelated model wins in 50 out of the 54 years
of data on which we have tested.
Comparing the anisotropic uncorrelated
model and the anisotropic correlated model we see that the anisotropic correlated model wins in
42 of the 54 years we have tested.

How might we actually quantify the significance of these differences?
It would be wrong to test the differences in the means between these models using a t-test
or bootstrap, since the different years are sampled from different distributions.
This is because the likelihood values in each year depend on the number of hurricanes in that year,
as well as the length of each track. The likelihood score for a poor model in a year with many long
hurricane tracks could easily be greater than that for a good model in a year in which there
are only a few short hurricane tracks.

One way to avoid this difficulty might be to normalise the
likelihood scores in each year using the number of tracks and the number
of points on the tracks, but that is hard to do in a fair way because of correlations between errors
along the tracks. As an alternative, we simply test whether 50 wins in 54 contests (and 42 wins in
54 contests) is significant, against the null hypothesis that winning is equally likely for either model.
This is then equivalent to testing whether 50 or more heads in 54 coin tosses would be significant
(and hence evidence for a biased coin). Using the CDF of the binomial distribution for 54 contests this
gives probabilities of over 0.99999999999 and over 0.99999 for our two tests.
We see that the differences between the three models are indeed highly significant.

Since our log-likelihood scores are all calculated out-of-sample we can conclude that the
anisotropic correlated model is the best of the three models at predicting the distribution of the
future positions of the hurricane track.
Thus, the contours of constant probability in the distribution of future positions of a
hurricane are significantly elliptical, and the principal axes of these ellipses are
significantly rotated relative to the mean track.

\subsection{Variance and correlation maps}

\begin{figure}[!htb]
  \begin{center}
    \scalebox{0.8}{\includegraphics{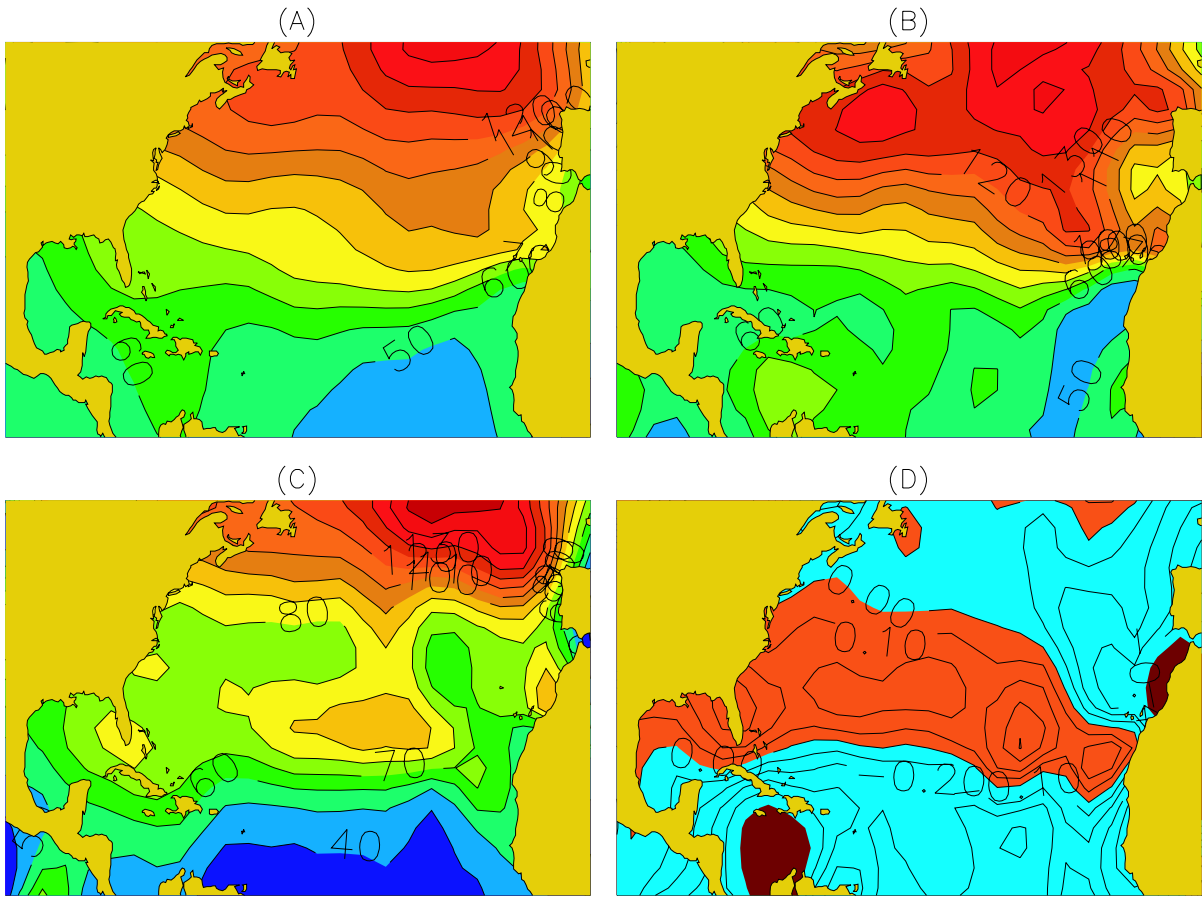}}
  \end{center}
  \caption{
Panel A shows the variance field for the isotropic model.
Panels B and C show the variance field for the anisotropic uncorrelated
model, in the along-mean-track and across-mean-track directions respectively,
and panel D shows the correlations from the anisotropic correlated model.
          }
  \label{f03}
\end{figure}

Having determined the optimum averaging length-scale within each model we can then calculate the
implied variance field. These variance fields can be considered as an optimally smoothed estimate
of the real variance field within the context of that particular model. Figure~\ref{f03} shows
these variance fields for two of the models.
In the top left panel we see the variance field for the isotropic model. The main feature
of this field is a steep gradient in variance from south to north, with the standard deviation
increasing from around 50km in the trade-wind region to over 100km in the westerlies.
In the top right and bottom left panels we see the variance fields for
the anisotropic correlated model, for along-mean-track and across-mean-track directions respectively.
The along-mean-track variances are slightly greater than the across-mean-track variances,
but both components show the strong
north-south gradient seen in the isotropic model variance field.

Finally in the lower right panel we show the correlation between the along
and across track deviations as determined by the optimal lengthscale in the anisotropic correlated
model. The correlations are everywhere rather weak, with values between -0.2 and 0.2.
This suggests that the ellipses describing the probability distribution for the deviations are
only slightly rotated relative to the unconditional mean tracks.

\begin{figure}[!htb]
  \begin{center}
    \scalebox{0.8}{\includegraphics{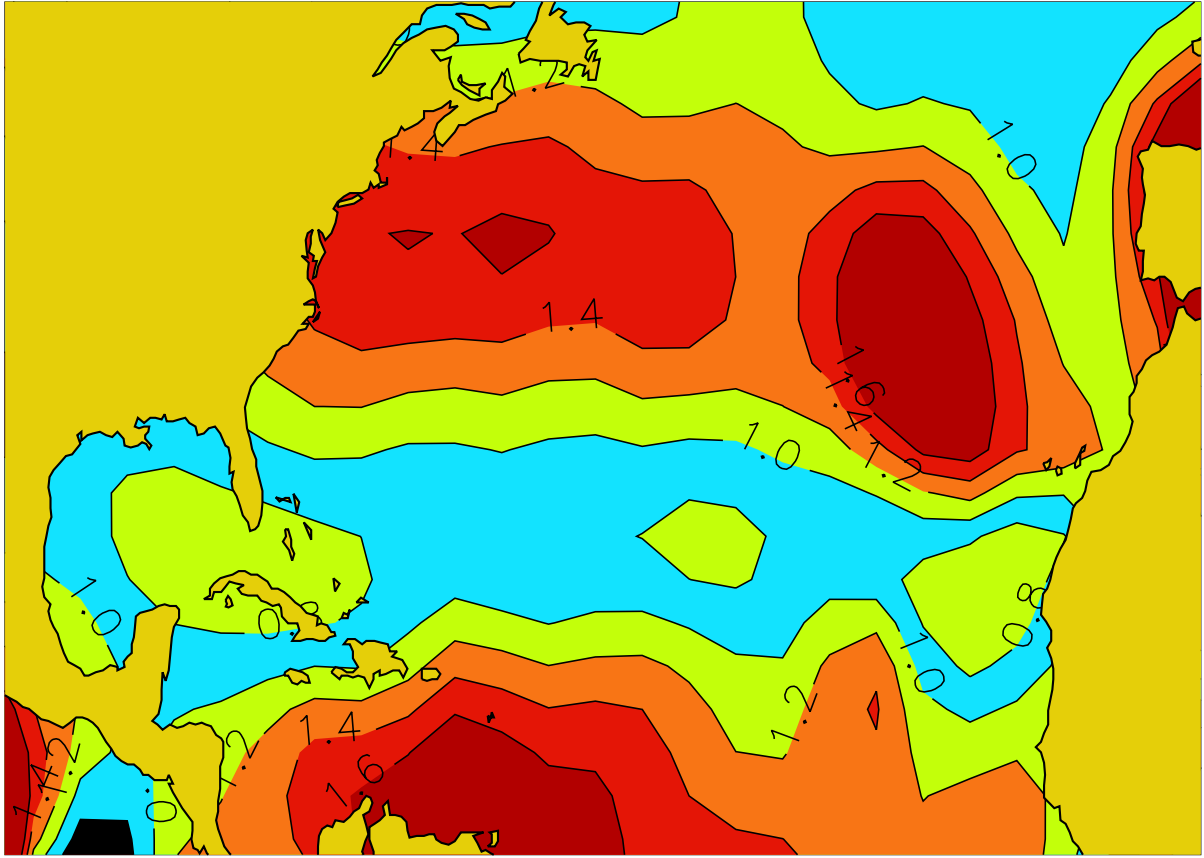}}
  \end{center}
  \caption{
The ratio of the along-mean-track variances to the across-mean-track variances
in the anisotropic uncorrelated model. This ratio indicates the extent to which
the contours of constant probability density deviate from circles and become
elliptical. A value near to 1 indicates the contours are nearly circular, while
a value above 1 indicates that the contours are elliptical, with the longest
axis along the direction of the unconditional mean tracks.
          }
  \label{f04}
\end{figure}

\begin{figure}[!htb]
  \begin{center}
    \scalebox{0.8}{\includegraphics{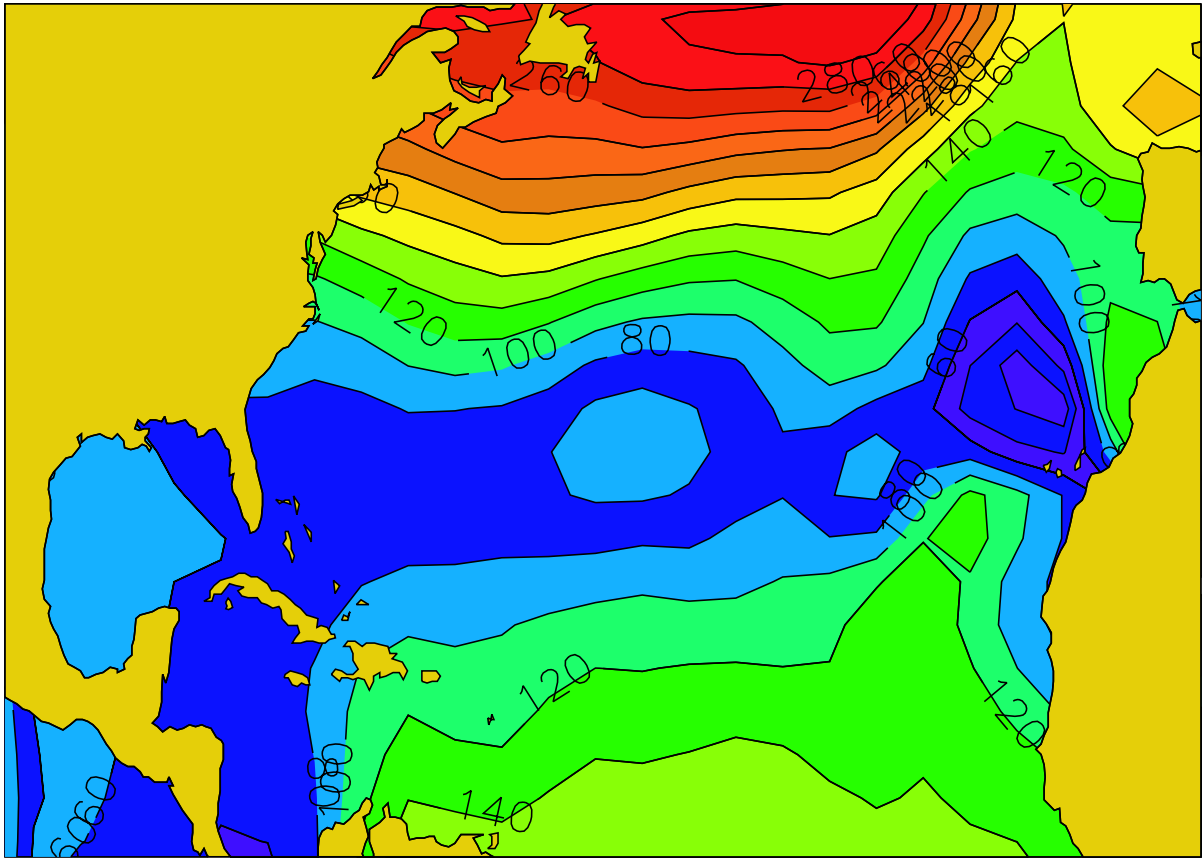}}
  \end{center}
  \caption{
The forward speed of the unconditional mean tracks.
          }
  \label{f05}
\end{figure}

In figure~\ref{f04} we show the ratio between the $u$ and $v$ variances from the anisotropic uncorrelated model.
This ratio is mostly positive, implying that errors in the along-mean-track direction are typically greater than
errors in the across-mean-track direction (as can already by seen by comparing the second and third panels
in figure~\ref{f03}).
There is also significant spatial structure in this ratio field,
which is perhaps related to the typical forward speeds of hurricanes in each region,
as shown in figure~\ref{f05}.

\subsection{Simulated tracks}

\begin{figure}[!htb]
  \begin{center}
    \scalebox{0.8}{\includegraphics{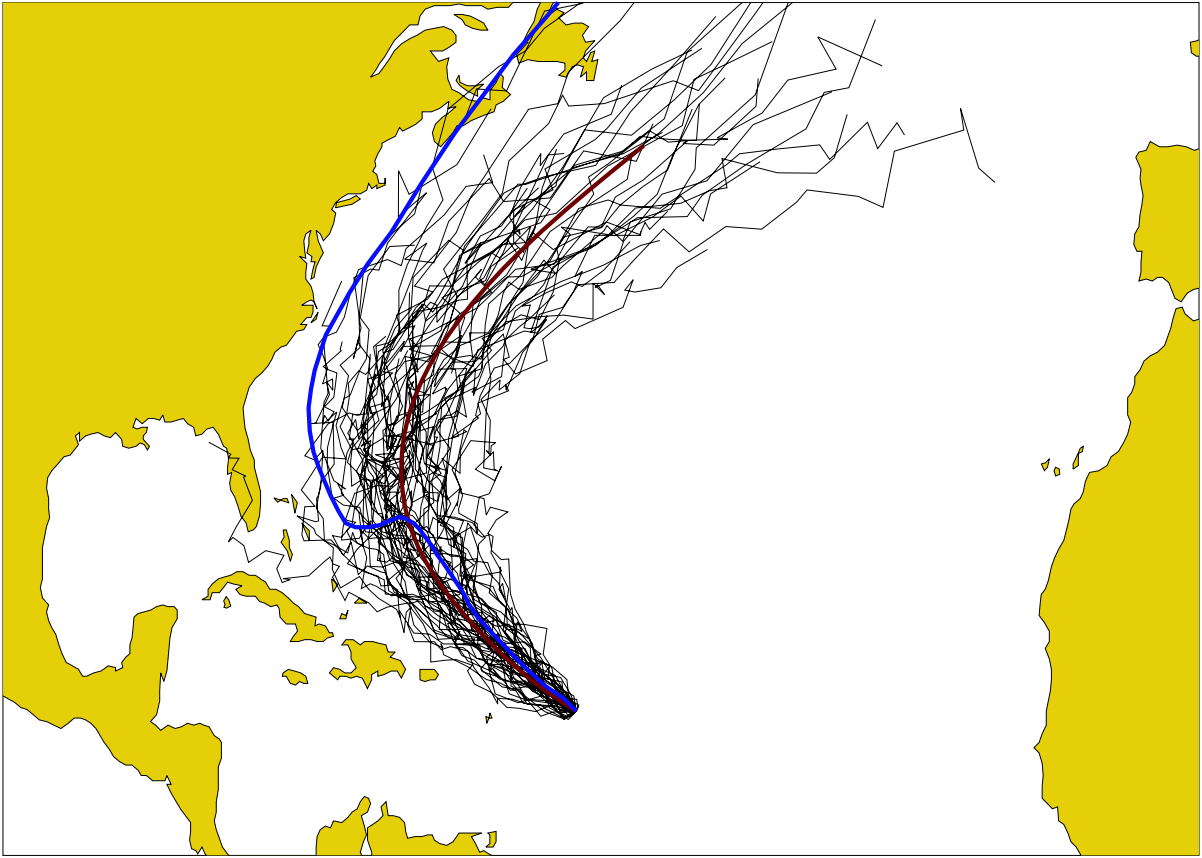}}
  \end{center}
  \caption{
The thin black curves show simulated hurricane tracks from the anisotropic correlated model,
all originating from the same point. The thick red curve shows the unconditional mean track from the same point,
and the blue curve shows an observed hurricane track.
          }
  \label{f06}
\end{figure}

Although we are still ignoring a number of features that are possibly rather important,
we now have the ingredients for a minimal stochastic model for hurricane tracks.
Figure~\ref{f06} shows some tracks generated from this model.
For comparison, one example of a real hurricane track is shown by the blue
line, and other examples are give in figure 1 in~\citet{hallj05a}.
The simulated tracks certainly show some of the features
of observed hurricane tracks: for instance, they move westward in the subtropical Atlantic
and eastward in mid-latitudes.
However, it seems clear
that these simulated tracks are less smooth than the observations. Presumably this is because
we are ignoring the autocorrelations between successive deviations along each track.
The incorporation of autocorrelations is the next stage in the development of the model.

\section{Summary}
\label{summary}

In a previous paper we have described a simple semi-parametric statistical model for the unconditional mean motion
of Atlantic hurricanes. In this paper we have investigated how to extend that model to include
the unconditional variance of the motion of the hurricanes too. The three models for the variance that we describe all
rely on a simple nearest-neighbour fitting technique,
with the length-scale that defines `near' being optimised using cross-validation.
Of these models, the simplest, which models the deviations from the unconditional
mean track to be variance-isotropic, performs the least well.
A model in which the deviations are modelled to be uncorrelated in directions along and across
the unconditional mean track,
but with different variances in these two directions, performs much better. Finally a model which
represents the deviations in these two directions as correlated performs better still, and is thus
the best of the models that we have tested.

The cost function we use to fit and compare our models is the out-of-sample likelihood. We propose
this cost function as a sensible and objective way to compare \emph{any} two hurricane track models.
For instance, the models of~\citet{vickery00} and~\citet{emanuel05} could be evaluated and compared with
our model using this cost function. As long as the models are fitted on the same underlying data,
and the cross-validation is performed correctly, this would be a fair way to compare the models.

Using the models described in this paper we can normalise the
observed deviations from the unconditional mean tracks
so that they become stationary in variance (as well as mean zero).
This will hopefully make them much easier to model using standard statistical methods
(which typically assume stationarity and constant variance).
The next stage of our modelling strategy is therefore to attempt to model
these standardised deviations using well-known time-series techniques such as
AR and ARMA, and thus incorporate memory into the
model.

\appendix

\section{Likelihood for the anisotropic correlated model}

For reference, we now give the likelihood for the anisotropic correlated model in terms
of the covariance $c$ rather than the correlation $r$.

The covariance matrix $\Sigma$ is given by

\begin{equation}
\Sigma
=
\left(
\begin{array}{cc}
  \sigma_u^2          & c \\
  c & \sigma_v^2          \\
\end{array}%
\right).
\end{equation}

The inverse of the covariance matrix is

\begin{equation}
\Sigma^{-1}
=
\frac{1}{\sigma_u^2 \sigma_v^2-c^2}
\left(
\begin{array}{cc}
  \sigma_v^2            & -c \\
  -c  & \sigma_u^2 \\
\end{array}%
\right)
\end{equation}

and
\begin{equation}
D
=\sigma_u^2 \sigma_v^2-c^2.
\end{equation}

This gives:
\begin{eqnarray}
(z-\mu)^T\Sigma^{-1}(z-\mu)
 &=& \frac{1}{\sigma_u^2\sigma_v^2-c^2}(u^2 \sigma_v^2+v^2 \sigma_u^2-2uvc)
\end{eqnarray}


The log-likelihood for $N$ data points is then:

\begin{eqnarray}
 \mbox{ln}f
         &=&\sum_{i=1}^{N} -\mbox{ln} [2\pi (\sigma_u^2 \sigma_v^2-c^2)^\frac{1}{2}]
           +\sum -\frac{1}{2(\sigma_u^2\sigma_v^2-c^2)}(u_i^2 \sigma_v^2+v_i^2 \sigma_u^2-2uvc)
\end{eqnarray}

Again we've suppressed the $i$ subscripts on $\sigma_u$, $\sigma_v$ and $c$ for simplicity of notation.

\section{Derivation of an advection-diffusion differential equation}
\label{fokker}

Purely for interest sake we now show that the models described above can be reformulated as an
advection-diffusion equation.

Given the stochastic differential equation in equation~\ref{isotropic}
one can apply some of the standard
machinery of mathematical physics. In particular, one can derive a partial differential equation
(usually known as the Fokker-Planck equation or the Kolmogorov equation)
that governs the evolution of a density field in space and time~\citep{gardiner85}.
In our case the density refers to the probability density of hurricanes.
If we write this density
as $f$, and ignore curvature effects, this gives:

\begin{equation}
 \frac{\partial f(\theta,\phi,t)}{\partial t}
  +\frac{\partial}{\partial \theta} [\mu_\theta f]
  +\frac{\partial}{\partial \phi} [\mu_\phi f]
  =\frac{1}{2}\frac{\partial^2}{\partial \theta^2} [\sigma^2 f]
  +\frac{1}{2}\frac{\partial^2}{\partial \phi^2} [\sigma^2 f]
  +\frac{\partial^2}{\partial \theta \partial \phi} [\rho \sigma^2 f]
\end{equation}

or, more succintly, as

\begin{equation}\label{fp}
 \frac{\partial f}{\partial t}
  +\nabla. (\mbox{\boldmath $\mu$} f)
  =\nabla^2 (\sigma f)
\end{equation}

Slightly more esoterically, there is also a partial differential equation for the evolution of the
probability density backwards in time (the backward Fokker-Planck or backwards Kolmogorov equation).
This can be written as:
\begin{equation}
 \frac{\partial f}{\partial t}
  =\mbox{\boldmath $\mu$}. \nabla f
  +\sigma \nabla^2 f
\end{equation}

We present these equations mainly for curiosity value. They don't seem to be particularly useful for solving
the practial problem of modelling hurricane risk since they don't generalise easily to the case where there is
memory along the trajectory. Also, we are ultimately interested in modelling the intensity along the track
and the damage caused by individual hurricanes, and these also don't fit into this framework. However, we find
the analogy suggested by equation~\ref{fp} reasonably interesting: the distribution of possible hurricanes is
advected by a mean flow field, including the effects of compression and expansion, and is diffused by a diffusive field.
The equation governing this behaviour is exactly the same as the equation that governs
the advection and diffusion of the density of a compressible fluid.

\section{Legal statement}

SJ was employed by RMS at the time that this article was written.

However, neither the research behind this article nor the writing
of this article were in the course of his employment, (where 'in
the course of their employment' is within the meaning of the
Copyright, Designs and Patents Act 1988, Section 11), nor were
they in the course of his normal duties, or in the course of
duties falling outside his normal duties but specifically assigned
to him (where 'in the course of his normal duties' and 'in the
course of duties falling outside his normal duties' are within the
meanings of the Patents Act 1977, Section 39). Furthermore the
article does not contain any proprietary information or trade
secrets of RMS. As a result, the authors are the owner of all the
intellectual property rights (including, but not limited to,
copyright, moral rights, design rights and rights to inventions)
associated with and arising from this article. The authors reserve
all these rights. No-one may reproduce, store or transmit, in any
form or by any means, any part of this article without the
authors' prior written permission. The moral rights of the authors
have been asserted.

The contents of this article reflect the authors' personal
opinions at the point in time at which this article was submitted
for publication. However, by the very nature of ongoing research,
they do not necessarily reflect the authors' current opinions. In
addition, they do not necessarily reflect the opinions of the
authors' employers.

\bibliography{timhall3}

\newpage

\newpage

\newpage

\newpage

\newpage

\newpage



\end{document}